\documentclass[11pt]{article}

\usepackage[T1]{fontenc}
\usepackage[utf8]{inputenc}
\usepackage{amsmath}
\RequirePackage[numbers,sort&compress]{natbib}
\usepackage{graphicx}
\usepackage{bm}
\usepackage[colorlinks=true,urlcolor=blue,linkcolor=blue,citecolor=red]{hyperref}
\renewcommand{\thefootnote}{\fnsymbol{footnote}}

\begin{document}

\begin{center}
{\Large \bf De Sitter-invariant approach to cosmology}
\vskip 0.5cm
{\bf A. V. Araujo,$^{a}$
D. F. L\'opez,$^{b}$
J. G. Pereira,$^{c,}$\footnote{Corresponding author: jg.pereira@unesp.br} 
J. R. Salazar$^{c}$}
\vskip 0.2cm
$^a${\it Departamento de Matem\'aticas,
Universidad Sergio\\ Arboleda, Bogot\'a, Colombia}
\vskip 0.2cm
$^b${\it Department of Mathematics and Statistics\\ 
Dalhousie University\\
Halifax, Canada} 
\vskip 0.2cm
$^c${\it Instituto de F\'{\i}sica Te\'orica \\ 
Universidade Estadual Paulista \\
S\~ao Paulo, Brazil}
\end{center}


\vskip 0.2cm
\centerline{\bf Abstract}
\begin{quote}
{\small The spacetime short-distance structure at the Planck scale is governed by the Planck length, usually interpreted as a three-dimensional Euclidian length. As such, it is not Lorentz invariant and clashes with Einstein's special relativity, which is thus unable to describe the Planck scale kinematics. The solution to this problem is twofold. First, one has to re-interpret the Planck length as a Lorentz invariant four-dimensional pseudo-length. Second, to comply with a non-vanishing cosmological term~$\Lambda$, one has to replace the standard Poincaré-invariant special relativity with the de Sitter-invariant special relativity. Since the Planck pseudo-length does not clash with the de Sitter-invariant special relativity, it provides a consistent description of the Planck scale kinematics in the presence of~$\Lambda$. Under the above replacement, general relativity changes to the de Sitter-invariant general relativity, in which~$\Lambda$ is constitutive. In this paper, the ensuing Friedmann equations are derived, and some implications for cosmology are explored and discussed.} 
\end{quote}
\renewcommand{\thefootnote}{\arabic{footnote}}
\setcounter{footnote}{0}

\section{Introduction}

A fundamental issue of spacetime physics is the lack of a special relativity theory suitable for describing the Planck scale kinematics. The problem is that the short-distance structure of spacetime at the Planck scale is governed by the Planck length, usually interpreted as a three-dimensional Euclidian length ($i, j, k, \dots = 1,2,3$)
\begin{equation}
l_P^2 \sim \delta_{ij}\, x^i x^j.
\end{equation}
Since this definition of Planck length is not Lorentz invariant, it clashes with Einstein's special relativity, which is thus unable to describe the Planck scale kinematics. For this reason, there is a widespread belief that Lorentz invariance should break down at the Planck scale.

However, such an interpretation of the Planck length has consistency problems. For example, in terms of fundamental constants of nature, the Planck length reads
\begin{equation}
l_P^2 = \frac{G \hbar}{c^3}.
\end{equation}
As the fundamental constants are Lorentz invariant, the Planck length should also be Lorentz invariant. To be Lorentz invariant, instead of being interpreted as a three-dimensional Euclidian length, the Planck length must be interpreted as a four-dimensional pseudo-length ($\alpha, \beta, \gamma, \dots = 0,1,2,3$)
\begin{equation}
l_P^2 \sim g_{\alpha \beta}\, x^\alpha x^\beta,
\label{LinePseudo}
\end{equation}
with $g_{\alpha \beta}$ a pseudo-Euclidian (or Lorentzian) metric. As we will see, more than the introduction of the Planck pseudo-length is needed to obtain the Planck scale kinematics in the presence of a non-vanishing cosmological term~$\Lambda$.

To begin with, we clarify that, whenever talking about locality, we mean the locality notion inherent to the strong equivalence principle. According to this principle, at every spacetime point in an arbitrary gravitational field, one can define a locally inertial frame in which inertial effects precisely compensate for gravitation. Consequently, inertial effects and gravitation go out of sight, and the laws of physics reduce locally to those of special relativity, as seen from an inertial frame~\cite{MTW}.
 
Note that this notion of locality differs from the usual geometric notion, in which a curved surface can be locally approximated by a flat surface. Although this approximation is correct, it is not a physical principle and has nothing to do with the strong equivalence principle~\cite{Synge}. Note also that since the cosmological term~$\Lambda$ represents neither gravitation nor inertial effect, it does not interfere with the strong equivalence principle and, consequently, with the notion of locality.

Minkowski and de Sitter are fundamental spacetimes in the sense that they can be algebraically constructed independently of Einstein's equation. They constitute non-gravitational backgrounds for constructing physical theories~\cite{GeoLivro}. General relativity, for example, can be constructed on any of them. In either case, gravitation will have the same dynamics: only the {\em local} kinematics will differ.

When general relativity is constructed on Minkowski, all solutions to the gravitational field equation are spacetimes that reduce locally to Minkowski, 
where spacetime's local kinematics occur. Since the kinematic group of Minkowski is the Poincaré group, spacetime's local kinematics will be governed by the standard Poincaré-invariant special relativity. In this case, the absence of gravity is represented by Minkowski spacetime.

When general relativity is constructed on de Sitter, all solutions to the gravitational field equation are spacetimes that reduce locally to de Sitter, where spacetime's local kinematics occur.\footnote{Spacetimes that do not reduce locally to Minkowski has been known for a long time and come under the name of Cartan geometry~\cite{1CartanGeoPRE,1wisePRE,HendrikCartan}.} Since the kinematic group of de Sitter is the de Sitter group, spacetime's local kinematics will be governed by the de Sitter-invariant special relativity~\cite{dSsr0,dSsr1}. In this case, the absence of gravity is represented by de Sitter spacetime.

Different from the standard Poincaré-invariant approach, where the cosmological term $\Lambda$ must be added by hand to the gravitational field equation, in the de Sitter-invariant approach, the cosmological term~$\Lambda$ shows up encoded in the spacetime's local kinematics and is defined as the sectional curvature of the non-gravitational background spacetime,
\begin{equation}
\Lambda \sim l^{-2},
\label{SectionalCurv}
\end{equation}
with $l$ the de Sitter pseudo-radius.\footnote{We see from Eq.~\eqref{constraint0} ahead that the de Sitter length parameter $l$ is a pseudo-length. Thus the name pseudo-radius.}
In the case of Poincaré-invariant general relativity, where spacetime reduces locally to the flat Minkowski spacetime, the cosmological term $\Lambda$ vanishes. Therefore, there is no room for a non-vanishing $\Lambda$ in standard general relativity. On the other hand, in the case of de Sitter-invariant general relativity, where spacetime reduces locally to de Sitter, the cosmological term~$\Lambda$ is non-vanishing, and the pseudo-radius $l$ is finite.
 
At the Planck scale, the de Sitter pseudo-radius~$l$ coincides with the Planck length~$l_P$. In this case, the Planck length $l_P$ represents the pseudo-radius of the background de Sitter spacetime, which will be a de Sitter space with the Planck cosmological term
\begin{equation}
\Lambda_P \sim l_P^{-2}.
\label{PlanckL}
\end{equation}
Since the Planck pseudo-length is Lorentz invariant, it does not clash with the de Sitter-invariant special relativity, which provides a consistent description of the Planck scale kinematics.

Once one reinterprets the Planck length as a Lorentz-invariant four-dimensional pseudo-length, the problem of the Planck scale kinematics and the problem of the universe's large-scale kinematics in the presence of~$\Lambda$ reduce to a single problem. The only difference is the value of the pseudo-radius: at the Planck scale, it coincides with the Planck pseudo-length~$l_P$, whereas at the universe's large scale, it is much larger than $l_P$. This picture points to a decaying huge primordial~$\Lambda$, which could drive inflation and then evolve to the small values we measure today.
It is important to remark that according to the de Sitter-invariant approach, physics remains Lorentz invariant at all energy scales, including the Planck scale.

By construction, the de Sitter-invariant approach to gravitation and cosmology constitutes a natural framework for the Planck scale physics and the universe's cosmological scale. Using this approach, we aim in this paper to obtain the de Sitter-invariant Friedmann equations and explore some of their implications for cosmology.

\section{Minkowski and de Sitter as quotient spaces}
\label{Quotient}

Spacetimes with constant sectional curvature are maximally symmetric because they carry the maximum number of Killing vectors. Flat Minkowski spacetime $M$ is the simplest one. Its kinematic group is the Poinca\-r\'e group ${\mathcal P} = {\mathcal L} \oslash {\mathcal T}$, the semi-direct product between Lorentz ${\mathcal L}$ and the translation group ${\mathcal T}$. Algebraically, it is defined as the quotient space
\[
M = {\mathcal P}/{\mathcal L}.
\]
The Lorentz subgroup is responsible for the isotropy around a given point of $M$, and the translation symmetry enforces this isotropy around all other points. In this case, homogeneity means that all points of Minkowski are equivalent under spacetime translations. One then says that Minkowski is {\em transitive} under translations, whose generators are written as
\begin{equation}
P_\nu = \delta^\alpha_\nu \partial_\alpha,
\label{transiM}
\end{equation}
with $\delta^\alpha_\rho$ the Killing vectors of spacetime translations.

The de Sitter space~$dS$ is also maximally symmetric, with the de Sitter group~$SO(1,4)$ as the kinematic group. Algebraically, it is defined as the quotient space
\[
dS = SO(1,4) / {\mathcal L}.
\]
Like Minkowski, the Lorentz subgroup is responsible for the isotropy around a given point of $dS$. The homogeneity, however, differs substantially. To find out the de Sitter homogeneity, let us recall that the de Sitter spacetime can be viewed as a hyperboloid embedded in the $(1+4)$-dimensional pseudo-Euclidean space with Cartesian coordinates $\chi^A$ ($A, B, C \dots = 0, \dots 4$) and metric
\[
\eta_{AB} = {\rm diag}\, (+1,-1,-1,-1,-1),
\]
inclusion whose points satisfy~\cite{HE}
\begin{equation}
\eta_{AB} \, \chi^A \chi^B = - \, l^2.
\label{constraint0}
\end{equation}
In terms of the coordinates~${\chi^A}$, the generators of infinitesimal de Sitter transformations are written as
\begin{equation}
L_{A B} \equiv \zeta^{\;C}_{AB} \, \frac{\partial}{\partial \chi^C}
\label{dsgene}
\end{equation}
where
\begin{equation}
\zeta^{\;C}_{AB} = \eta_{AD} \chi^D \, \delta^C_B -
\eta_{BD} \chi^D \, \delta^C_A \,
\end{equation}
are the associated Killing vectors.

Upon writing these generators in terms of the four-dimensional stereographic coordinates $\{x^\mu\}$, there are two possible parameterizations: one appropriate for small values of $\Lambda$ and another appropriate for large values of $\Lambda$, as compared with the Planck cosmological term~\cite{jgp}. Since our interest in this paper is to study the present-day universe, we will consider only the parameterization appropriate for small values of~$\Lambda$. In this case, the ten de Sitter generators~(\ref{dsgene}) are written as~\cite{gursey}
\begin{equation}
L_{\mu \nu} = \zeta^{\,\alpha}_{\mu \nu} \, \partial_\alpha \qquad \mbox{and} \qquad
{\Pi}_\nu \equiv \frac{L_{4 \nu}}{l} = \xi^\alpha_\nu \partial_\alpha,
\label{dSgene4dim}
\end{equation}
where $L_{\mu \nu}$ represent the Lorentz generators and ${\Pi}_\nu$ stand for the so-called de Sitter ``translation'' generators, with $\zeta^{\,\alpha}_{\mu \nu}$ and $\xi^\alpha_\nu$ the corresponding Killing vectors.\footnote{The generators $\Pi_\nu$ are not really translations but rotations in the planes $(4,\nu)$. Hence, the quotation marks.} In contrast to Minkowski, whose points are equivalent under spacetime translations, all points of de Sitter are equivalent under de Sitter ``translations.'' One then says that de Sitter is transitive under de Sitter ``translations.''

In stereographic coordinates, the Killing vectors of the de Sitter ``translations'' split in the form~\cite{ccc}
\begin{equation}
\xi^\alpha_\nu =
\delta^\alpha_\nu - \frac{1}{4 l^2} \, \vartheta^\alpha_\nu,
\label{dsKilling3Bis}
\end{equation}
where
\begin{equation}
\delta^\alpha_\nu \quad \mbox{and} \quad \vartheta^\alpha_\nu = 2 \eta_{\nu \rho} \, x^\rho x^\alpha -
\sigma^2 \delta^\alpha_\nu
\end{equation}
are, respectively, the Killing vectors of translations and proper conformal transformations. Consequently, the de Sitter ``translation'' generators can be recast in the form
\begin{equation}
{\Pi}_\nu = {P}_\nu - \frac{1}{4 l^2}\, {K}_\nu,
\label{pi3}
\end{equation}
where
\begin{equation}
{P}_\nu = \delta^\alpha_\nu \, \partial_\alpha \quad \mbox{and} \quad
{K}_\nu = \vartheta^\alpha_\nu \, \partial_\alpha
\label{TransGenerators}
\end{equation}
are, respectively, the translation and proper conformal gene\-ra\-tors~\cite{coleman}. Equa\-tions~\eqref{transiM} and \eqref{pi3} show that, {\em whereas Minkowski is transitive under translations, the de Sitter spacetime is transitive under a combination of translations and proper conformal transformations}~\cite{dSgeod}.

\section{De Sitter-invariant general relativity}
\label{dSmGR}

This section presents the schematic procedure to obtain the de Sitter-invariant Einstein's equation. For comparison, we first review how the standard Einstein's equation is usually obtained.

\subsection{Poincar\'{e}-invariant Einstein's equation}

In the case of standard general relativity, all solutions to the gravitational field equations are spacetimes that reduce locally to Minkowski. Since Minkowski is transitive under translations, a diffeomorphism in these spacetimes is defined as a local translation
\begin{equation}
\delta_P x^\mu = \delta^\mu_\alpha \, \epsilon^\alpha(x),
\label{OrDiff}
\end{equation}
with $\delta^\mu_\alpha$ the Killing vectors of translations. From Noether's theorem, the invariance of the source Lagrangian under the diffeomorphism~\eqref{OrDiff} yields the energy-momentum covariant conservation law
\begin{equation}
\nabla_\nu T^{\mu \nu} = 0 \qquad \mbox{with} \qquad T^{\mu \nu} = \delta^{\mu}_{\alpha} \, T^{\alpha \nu}.
\end{equation}
Variation of the gravitational plus source actions under the diffeomorphism~\eqref{OrDiff} yields the standard Einstein's equation \begin{equation}
G^{\mu \nu} \equiv R^{\mu \nu} - {\textstyle{\frac{1}{2}}} g^{\mu \nu} R = - \frac{8 \pi G}{c^4} \,T^{\mu \nu}.
\label{OrdiEinstein}
\end{equation}

\subsection{De Sitter-invariant Einstein's equation}

In the case of the de Sitter-invariant general relativity, all solutions to the gravitational field equation are spacetimes that reduce locally to de Sitter. Since de Sitter is transitive under de Sitter ``translations,'' a diffeomorphism in these spacetimes is defined as~\cite{dSgeod}
\begin{equation}
\delta_{\Pi} x^\mu = \xi^{\mu}_{\alpha} \, \epsilon^{\alpha}(x),
\label{dStrans}
\end{equation}
with $\xi^{\mu}_{\alpha}$ the Killing vectors~\eqref{dsKilling3Bis} of the de Sitter ``translations.'' From Noether's theorem, the invariance of the source Lagrangian under the diffeomorphism~\eqref{dStrans} yields the covariant conservation law
\begin{equation}
\nabla_\nu \Pi^{\mu \nu} = 0 \qquad \mbox{with} \qquad \Pi^{\mu \nu} = \xi^{\mu}_{\alpha} \, T^{\alpha \nu}.
\label{dSconservation4}
\end{equation}
Variation of the gravitational plus source actions under the diffeomorphism~\eqref{dStrans} yields the de Sitter-invariant Einstein's equation~\cite{dSde}
\begin{equation}
{\mathcal G}^{\mu \nu} \equiv {\mathcal R}^{\mu \nu} - {\textstyle{\frac{1}{2}}} g^{\mu \nu} {\mathcal R} = - \frac{8 \pi G}{c^4} \, \Pi^{\mu \nu},
\label{NewModiEinstein}
\end{equation}
with ${\mathcal R}^\alpha{}_{\beta \mu \nu}$ the curvature Riemann tensor. In locally de Sitter spacetimes, the metric can be locally decomposed in the form
\[
g_{\mu \nu} \doteq \hat g_{\mu\nu} + h_{\mu \nu},
\]
where $\hat g_{\mu\nu}$ is the background de Sitter metric and $ h_{\mu \nu}$ is a tensor that represents the gravitational field. In this case, the Riemann tensor  ${\mathcal R}^\alpha{}_{\beta \mu \nu}$ computed out of the metric $g_{\mu \nu}$ includes both the kinematic curvature of the background de Sitter spacetime and the dynamic curvature of general relativity. 

\subsection{The source of the cosmological term}
\label{dSsource}

Substituting the Killing vectors~\eqref{dsKilling3Bis} in the source current~$\Pi^{\mu \nu}$, it splits in the form
\begin{equation}
\Pi^{\mu \nu} = T^{\mu \nu} - ({1}/{4 l^2})\, K^{\mu \nu},
\label{Pi=T-K}
\end{equation}
where
\begin{equation}
T^{\mu \nu} = \delta^\mu_\alpha \, T^{\alpha \nu}\qquad \mbox{and} \qquad K^{\mu \nu} = \vartheta^\mu_\alpha \, T^{\alpha \nu}
\label{T&K}
\end{equation}
are, respectively, the energy-momentum and the proper conformal currents~\cite{ConfCurrent}.
Analogously to the source decomposition, the Einstein tensor ${\mathcal G}^{\mu \nu}$ splits in the form
\begin{equation}
{\mathcal G}^{\mu \nu} = G^{\mu \nu} - \hat{G}^{\mu \nu},
\end{equation}
where ${G}^{\mu \nu}$ is general relativity's Einstein tensor and $\hat{G}^{\mu \nu}$ is the Einstein tensor of the local de Sitter spacetime. In stereographic coordinates, therefore, the de Sitter-invariant Einstein's equation~(\ref{NewModiEinstein}) assumes the form
\begin{equation}
\big({R}^{\mu \nu} - {\textstyle{\frac{1}{2}}} g^{\mu \nu} {R}\big) -
\big(\hat{R}^{\mu \nu} - {\textstyle{\frac{1}{2}}} g^{\mu \nu} \hat{R}\big)
= - \frac{8 \pi G}{c^4} \Big[T^{\mu \nu} - (1/4l^2)\, K^{\mu \nu} \Big].
\label{NewEinstein2}
\end{equation}
According to this equation, in spacetimes that reduce locally to de Sitter, any source gives rise to an energy momentum and a proper conformal current. The energy-momentum current $T^{\mu \nu}$ keeps its role as the source of general relativity's dynamic curvature. In contrast, the proper conformal current $K^{\mu \nu}$ appears as the source of the kinematic curvature of the local de Sitter spacetime~\cite{SourceLambda}.

Note that the energy-momentum current is no longer covariantly conserved. What is conserved now is the combination of energy-momentum and proper conformal currents. Note also that whereas the gravitational part of the equation is essentially dynamic, the conformal part is purely algebraic. This difference stems from the non-propagating character of the cosmological term $\Lambda$.
Taking the trace of Eq.~\eqref{NewEinstein2} and identifying $
{G}^\mu{}_\mu = - R$ and $\hat{G}^\mu{}_\mu = - \Lambda$, it reduces to
\begin{equation}
R - \Lambda = \frac{8 \pi G}{c^4} \Big[ T^\mu{}_\mu - ({1}/{4 l^2})\, K^\mu{}_\mu \Big].
\label{NewEinsteinTrace}
\end{equation}

\subsection{Negative pressure from ordinary matter}

We consider now a perfect fluid whose energy-momentum tensor in co-moving coordinates is written as
\begin{equation}
T^\mu{}_\nu = {\rm diag} \left(\varepsilon_m, - p_m, - p_m, - p_m \right),
\label{CC1}
\end{equation}
where $\varepsilon_m$ and $p_m$ are the matter-energy density and pressure, respectively. Its trace has the form
\begin{equation}
T^{\mu}{}_{\mu} \equiv \delta^\mu_\alpha T^\alpha{}_\mu = \varepsilon_m - 3 p_m.
\label{TraceEM}
\end{equation}
On the other hand, the trace of the proper conformal current is
\begin{equation}
K^{\mu}{}_\mu = \vartheta^\mu_\alpha T^\alpha{}_\mu \equiv
\big(2 \eta_{\alpha \rho} x^\rho x^\mu - \sigma^2 \delta^\mu_\alpha \big) T^\alpha{}_\mu.
\label{Ktrace}
\end{equation}

Since the space section of the universe is assumed to be a homogeneous space in which all points are equivalent, one can compute the trace at any point~\cite{Wcosmo}. For the sake of simplicity, one usually chooses the point $x^i = 0$, which yields
$
K^\mu{}_\mu = c^2 t^2 \big(\varepsilon_m + 3 p_m \big)$.
We can then write
\begin{equation}
\frac{K^\mu{}_\mu}{4 l^2} = \gamma(t) \big(\varepsilon_m + 3 p_m \big),
\label{KtraceBis}
\end{equation}
where
\begin{equation}
\gamma(t) = {c^2 t^2}/{4 l^2}
\label{gamma}
\end{equation}
is a time-dependent dimensionless parameter. Considering that the proper conformal current is the source of dark energy, one can identify
\begin{equation}
\varepsilon_\Lambda \equiv \gamma(t) \, \varepsilon_m \quad \mbox{and} \quad
p_\Lambda \equiv \gamma(t)\, p_m.
\label{Lenr&pre}
\end{equation}
In this case, the trace \eqref{KtraceBis} of the proper conformal current reads
\begin{equation}
\frac{K^\mu{}_\mu}{4 l^2} = \varepsilon_\Lambda + 3 p_\Lambda.
\label{KtraceBis2}
\end{equation}
Using Eqs.~\eqref{TraceEM} and \eqref{KtraceBis2}, the field equation~\eqref{NewEinsteinTrace} can be recast in the form
\begin{equation}
{R} - \Lambda = \frac{8 \pi G}{c^4} \Big[\big(\varepsilon_m - 3 p_m \big) -
\big(\varepsilon_\Lambda + 3 p_\Lambda \big) \Big].
\label{NewEinsteinTrace2}
\end{equation}

On the other hand, as can be seen from~\eqref{Lenr&pre}, the energy densities $\varepsilon_\Lambda$ and $\varepsilon_m$ differ by the same coefficient as $p_\Lambda$ differs from $p_m$. Consequently, $p_m$ and $p_\Lambda$ satisfy the same equation of state,
\begin{equation}
p_m = w\, \varepsilon_m \qquad \mbox{and} \qquad p_\Lambda = w\, \varepsilon_\Lambda,
\label{EquState}
\end{equation}
with $w$ a numerical constant. Considering that the same ordinary matter is the source of both gravitation and dark energy, this is an expected result.
Comparing the trace~\eqref{TraceEM} of the energy-momentum current with the trace~\eqref{KtraceBis2} of the proper conformal current, we see that, even though $p_m$ and $p_\Lambda$ satisfy the same equation of state, they naturally enter the field equation~\eqref{NewEinsteinTrace2} with opposite signs concerning $\varepsilon_m$ and $\varepsilon_\Lambda$, respectively. This difference is due uniquely to the mathematical intricacies of the proper conformal current, the source of the background de Sitter spacetime. Therefore, in the de Sitter-invariant approach to cosmology, the source of dark energy does not need to satisfy an exotic equation of state to produce a repulsive interaction.

In the usual case of Poincar\'{e}-invariant general relativity, because the cosmo\-logical term $\Lambda$ is constant, the dark energy density $\varepsilon_\Lambda$ is uniformly distributed across space, independently of the matter energy-density $\varepsilon_m$ distribution. However, in the de Sitter-invariant general relativity, these energy densities are not independent but satisfy the constraint~\eqref{Lenr&pre}, a property inherited from the dependence of the proper conformal current~$K^{\mu \nu}$ on the energy-momentum current~$T^{\mu \nu}$.

\subsection{A hierarchy of kinematics and gravity theories}

One can establish a natural hierarchy of kinematics by using the In\"{o}n\"{u}-Wigner process of Lie groups expansion and contraction~\cite{gil,inonu}. At the bottom of the hierarchy stands the Galilei-invariant special relativity, which governs Newtonian gravity kinematics. The Poincar\'{e}-invariant Einstein's special relativity represents a generalization of Galilei relativity for velocities near the velocity of light.

\begin{figure}[hbtp]
\begin{center}
\scalebox{0.45}{\includegraphics{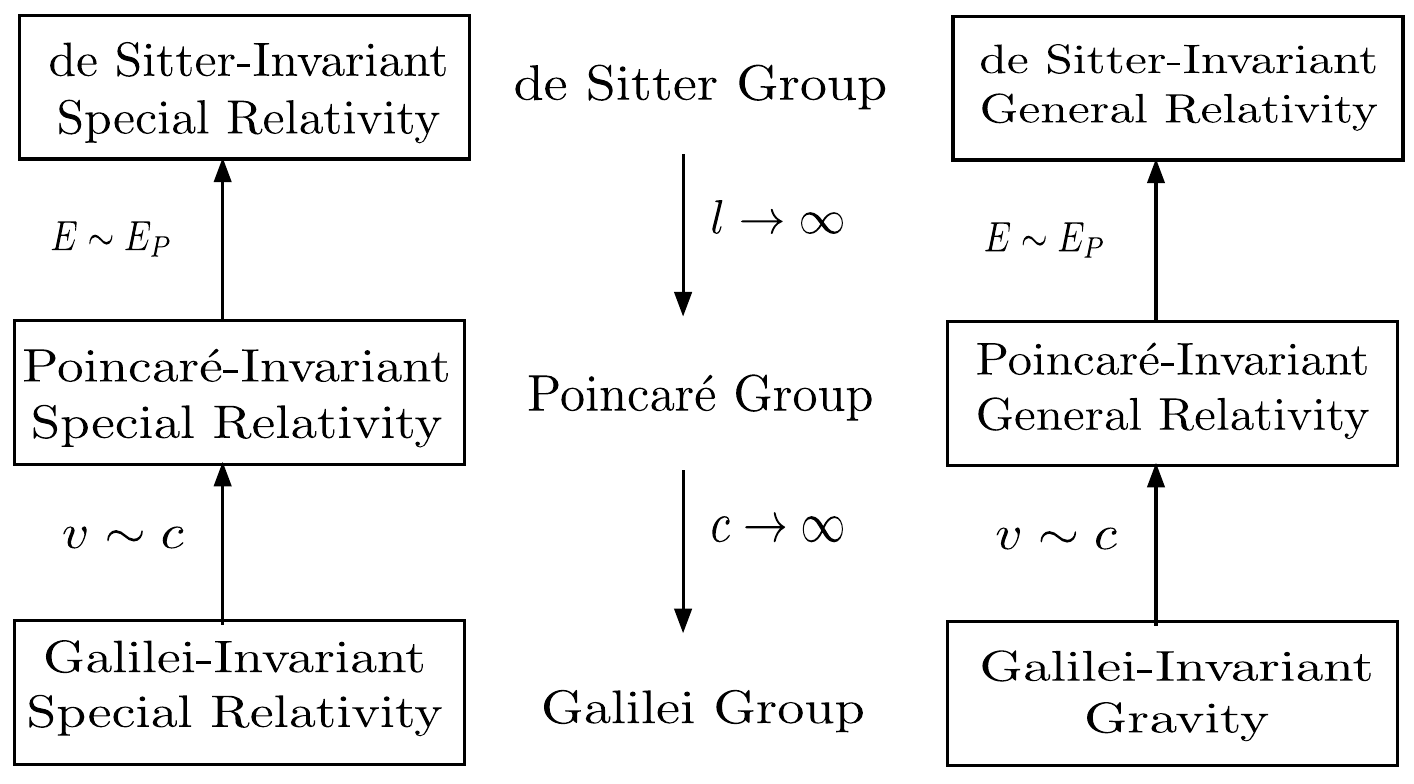}}
\caption{\em Pictorial view of the hierarchy of kinematics and gravity theories.}
\label{PereFig1}
\end{center}
\end{figure}
Accordingly, this theory gives rise to deviations concerning Galilei's relativity for velocities comparable to the velocity of light.
Similarly, the de Sitter-invariant special relativity may be interpreted as a generalization of Poincar\'e-invariant Einstein's special relativity for energies comparable to the Planck energy. Accordingly, this theory is expected to produce deviations concerning the Poincar\'{e}-invariant special relativity for energies comparable to the Planck energy and at the universe's large scale.

Conversely, the contraction limit $l \to \infty$ reduces the de Sitter-invariant special relativity to the Poincar\'{e}-invariant Einstein's special relativity. This theory reduces to the Galilei special relativity under the further contraction limit $c \to \infty$. Note that, to each special relativity, there corresponds a gravitational theory whose spacetime's local kinematics is governed by that special relativity. Figure~\ref{PereFig1} shows a pictorial view of the hierarchy of kinematics and the corresponding gravitational theories.

\section{De Sitter-invariant Friedmann equations}

\subsection{FLRW metric in locally de Sitter spacetimes}

The FLRW metric is constructed to comply with the cosmological principle, according to which the space section of the universe at large enough scales is assumed to be isotropic and homogeneous. There are only three possibilities: the space section can be Euclidean, spheric, or hyperbolic. Using the standard procedure, the FLRW metric is written as
\begin{equation}
ds^2 = c^2 dt^2 - a^2 \, \gamma_{ij}\, dx^i dx^j
\label{FRW1}
\end{equation}
where
$a = a(t)$ is the cosmic scale factor, and
\begin{equation}
\gamma_{i j} = \delta_{ij} + \frac{k\, x_i x_j}{1 - k(x_l x^l)},
\end{equation}
with $k$ the curvature parameter. For $k = 0, +1, -1$, the universe space section will be Euclidian, spheric, or hyperbolic, respectively.

The non-vanishing components of the Levi-Civita connection of the metric~\eqref{FRW1} are
\begin{equation}
\Gamma^0{}_{ij} = (a \dot{a}/c)\; \gamma_{ij}, \quad \Gamma^i{}_{0 j} = (\dot{a}/ac)\; \delta^i_j,
\quad \Gamma^i{}_{jk} = - x^i \delta_{jk},
\label{connections0}
\end{equation}
where a dot represents a derivative concerning the cosmic time $t$.

\subsection{Noether continuity equations}

Recalling that $\Pi^\mu{}_\nu$ denotes the symmetric part of the tensor, the zero-component of the covariant conservation law (\ref{dSconservation4}) reads
\begin{equation}
\nabla_{\mu}\Pi^{\mu}{}_{0} \equiv \partial_{\mu} \Pi^{\mu}{}_{0} +
\Gamma^{\mu}{}_{\rho \mu} \Pi^{\rho}{}_{0} -
\Gamma^{\rho}{}_{0 \mu} \Pi^{\mu}{}_{\rho} = 0.
\label{CE1}
\end{equation}
Separating the time and space components and noting that $\Pi^j{}_0 = \Pi_0{}^j$ vanishes due to the homogeneity of the universe space-section, we obtain
\begin{equation}
\partial_0 \Pi^0{}_0 + \Gamma^j{}_{0 j} \Pi^0{}_0 - \Gamma^i{}_{0 j} \Pi^j{}_i = 0.
\end{equation}
Substituting the connections~\eqref{connections0} computed at $x^i = 0$, we get
\begin{equation}
\partial_0 \Pi^0{}_0 + 3 \frac{\dot{a}}{a c} \, \Pi^0{}_0 -
\frac{\dot{a}}{a c} \, \Pi^j{}_j = 0.
\label{CE2}
\end{equation}
Using Eqs.~\eqref{Pi=T-K} and \eqref{T&K}, the components of the source current, computed at $x^i = 0$, are found to be
\begin{equation}
\Pi^0{}_0 = \left(\varepsilon_m - \varepsilon_\Lambda \right)
\label{1Pi's}
\end{equation}
and
\begin{equation}
\Pi^j{}_j = - 3 \left(p_m + p_\Lambda \right).
\label{2Pi's}
\end{equation}
Using the identifications~\eqref{Lenr&pre}, the conservation law~\eqref{CE2} can be rewritten in the form
\begin{equation}
\frac{d}{dt} \big(\varepsilon_m - \varepsilon_\Lambda \big) +
3 \frac{\dot{a}}{a} \, \big(\varepsilon_m - \varepsilon_\Lambda \big) +
3 \frac{\dot{a}}{a} \, \big(p_m + p_\Lambda \big) = 0.
\label{CE4}
\end{equation}

Assuming that the energy densities depend on the cosmic time $t$ through the scale factor $a = a(t)$, and using the equations of state~\eqref{EquState}, the conservation law~\eqref{CE4} becomes
\begin{equation}
\frac{d}{da} \Big[ a^3 \big(\varepsilon_m - \varepsilon_\Lambda\big) \Big] +
3 a^2 w (\varepsilon_m + \varepsilon_\Lambda) = 0.
\label{CE5}
\end{equation}
Its solution is
\begin{equation}
\varepsilon_m - \varepsilon_\Lambda = \beta \left( a^{- 3 - 3w} -
 a^{- 3 + 3 w} \right)
\label{Solution}
\end{equation}
with $\beta$ an integration constant. In the contraction limit $l \to \infty$, the continuity equation \eqref{CE5} reduces to the usual expression of locally-Minkowski spacetimes
\begin{equation}
\frac{d}{da} \big( a^3 \varepsilon_m \big) + 3 a^2 w \, \varepsilon_m = 0,
\label{CE5GR}
\end{equation}
whose solution is
\begin{equation}
\varepsilon_m = \beta \, a^{-3 - 3 w}.
\end{equation}

\subsection{Friedmann equations}

Using the Levi-Civita components~\eqref{connections0}, the non-vanishing components of the Ricci tensor computed at $x^i = 0$ are found to be
\begin{eqnarray}
&& {\mathcal R}_{00} = 3 \frac{\ddot{a}}{a c^2} \label{Ricci00} \\
&& {\mathcal R}_{ij} = -\frac{1}{c^2}\big(\ddot{a} a + 2 \dot{a}^2 - 2 c^2 \big) \delta_{ij}. \label{Riccijj}
\end{eqnarray}
The corresponding scalar curvature is
\begin{equation}
{\mathcal R} = \frac{6}{c^2} \left(\frac{\ddot{a}}{a} +
\frac{\dot{a}^2}{a^2} - \frac{ c^2}{a^2} \right).
\label{RicciScalar}
\end{equation}
Using these tensors in the de Sitter-invariant Einstein equation~\eqref{NewModiEinstein}, we obtain the de Sitter-invariant Friedmann equations
\begin{equation}
H^2 = \frac{8 \pi G}{3 c^2} (\varepsilon_m - \varepsilon_\Lambda) -
\frac{k c^2}{a^2}
\label{dSFried1}
\end{equation}
and
\begin{equation}
\frac{\ddot{a}}{a} = - \frac{4 \pi G}{3 c^2} \Big[ \left(\varepsilon_m - \varepsilon_\Lambda\right)+
3 \left(p_m + p_\Lambda\right) \Big],
\label{dSFried2}
\end{equation}
with $H = \dot{a}/a$ the Hubble parameter.
Similarly to the ordinary Friedmann equations, taking the time derivative of equation~\eqref{dSFried1}, and using~\eqref{dSFried2} to eliminate $\ddot{a}$, one obtains the conservation law~\eqref{CE4}.

The opposed signs of $\varepsilon_m$ and $\varepsilon_\Lambda$ come from their different physical effects: whereas the matter's energy density represents an attractive effect, the dark energy density represents a repulsive effect. Since the Friedmann equations are dynamic, it is natural that $\varepsilon_m$ and $\varepsilon_\Lambda$ enter the equations with opposite signs. In the de Sitter-invariant approach to cosmology, the difference $\varepsilon_m - \varepsilon_\Lambda$ is the fundamental variable for describing the universe dynamics. In other words, the balance between $\varepsilon_m$ and $\varepsilon_\Lambda$ determines how the universe evolves along with the cosmic time.\footnote{The negative sign of $\varepsilon_\Lambda$ is inherent to the de Sitter group. If the spacetime's local kinematics were governed by the anti-de Sitter group, the sign of $\varepsilon_\Lambda$ would be positive, leading to an unstable universe.} Note that despite $\varepsilon_m$ and $\varepsilon_\Lambda$ can evolve along with the cosmic time, they do it in such a way that the difference $\varepsilon_m - \varepsilon_\Lambda$ satisfies the Noether continuity equation~\eqref{CE4}.

\section{Some physical implications}

\subsection{The topology of the universe}
\label{topo}

Considering the critical energy density ${3 H^2 c^2}/{8 \pi G} = \varepsilon_c$, the Friedmann equation~\eqref{dSFried1} can be rewritten in the form
\begin{equation}
1 = \Omega_m - \Omega_{\Lambda} - \frac{k c^2}{H^2 {a}^2},
\label{dSFried1bis2}
\end{equation}
where
\begin{equation}
\Omega_m \equiv \frac{\varepsilon_m}{\varepsilon_c} = \frac{8 \pi G}{3H^2 c^2} \varepsilon_m \quad
\mbox{and} \quad \Omega_\Lambda \equiv \frac{\varepsilon_\Lambda}{\varepsilon_c} = \frac{8 \pi G}{3H^2 c^2} \varepsilon_\Lambda
\label{OmegaS}
\end{equation}
are, respectively, the matter and the dark-energy density parameters.

An essential feature of the de Sitter-invariant approach is that the energy densities $\varepsilon_m$ and $\varepsilon_\Lambda$ are not independent, a property inherited from the dependence of the proper conformal current $K^{\mu \nu}$ on the energy-momentum tensor $T^{\mu \nu}$, as can be seen from Eq.~\eqref{T&K}. This dependence establishes the constraint
\begin{equation}
\varepsilon_\Lambda = \gamma(t) \, \varepsilon_m,
\end{equation}
with $\gamma(t)$ given by Eq.~\eqref{gamma}. Of course, the density parameters $\Omega_\Lambda$ and $\Omega_m$ are also related through
\begin{equation}
\Omega_\Lambda = \gamma(t) \, \Omega_m.
\end{equation}
Using the Planck Collaboration values~\cite{PlanckResults}
\begin{equation}
\Omega_\Lambda \simeq 0.69 \qquad \mbox{and} \qquad \Omega_m \simeq 0.31,
\label{ObsOmegas}
\end{equation}
the present-day value of $\gamma(t)$ is
\begin{equation}
\gamma(t_0) \equiv \frac{\Omega_\Lambda}{\Omega_m} \simeq 2.2.
\label{gammazero}
\end{equation}
Consequently, the present-day value of the last term on the right-hand side of Eq.~\eqref{dSFried1bis2}, which represents effects coming from the universe topology, is
\begin{equation}
- \frac{k c^2}{H^2 {a}^2} \simeq 1.38.
\end{equation}
This relation implies that $k = -1$, pointing to a universe with hyperbolic space sections.

Note that the values~\eqref{ObsOmegas} yields the relation
\begin{equation}
\Omega_m + \Omega_\Lambda = 1.
\label{inventory}
\end{equation}
According to the Poincar\'{e}-invariant $\Lambda$CDM model, the above relation represents a dynamic equation and hints at a universe with a flat ($k=0$) space section. Unlike the density parameters $\Omega_m$ and $\Omega_\Lambda$, which represent natural constituents of the universe, the last term on the right-hand side of Eq.~\eqref{dSFried1bis2} does not represent a constituent of the universe. Even though it contributes to the universe dynamics, as far as the universe's inventory is concerned, only $\Omega_m$ and $\Omega_\Lambda$ must be considered. Since it is irrelevant to the universe's inventory whether their effects are attractive or repulsive, in the de Sitter-invariant approach, equation~\eqref{inventory} is just an algebraic description of the universe's inventory without any dynamical meaning.

\subsection{The current value of $\Lambda$}

Let us pick up the conformal part of the field equation~\eqref{NewEinsteinTrace}, which gives $\Lambda$ in terms of the trace of the proper conformal current:
\begin{equation}
\Lambda = \frac{8 \pi G}{c^4}\, \frac{K^\mu{}_\mu}{4 l^2}.
\label{NewEinsteinTraceBis}
\end{equation}
Considering that the energy content of the present-day universe can be assumed to be preponderantly in the form of dust $(p_m = 0)$, the trace of the proper conformal current computed at $x^i = 0$ is
\begin{equation}
K^\mu{}_\mu = \gamma(t)\, \varepsilon_m,
\end{equation}
where $\gamma(t)$ is a dimensionless parameter relating the dark and matter energy densities: $\gamma(t) = \varepsilon_\Lambda/ \varepsilon_m$. Equation~\eqref{NewEinsteinTraceBis} can then be rewritten in the form
\begin{equation}
\Lambda = \frac{8 \pi G}{c^4}\, \gamma(t)\, \varepsilon_m.
\label{530}
\end{equation}
Since $\gamma(t) = \varepsilon_\Lambda/\varepsilon_m$ is currently of order unity,\footnote{In the Poincar\'{e}-invariant approach, since $\varepsilon_\Lambda$ is constant and $\varepsilon_m$ changes along the cosmic time, the condition $\varepsilon_\Lambda/\varepsilon_m \sim 1$ can only be interpreted as a coincidence~\cite{Coincidence}. However, in the de Sitter-invariant approach, where both $\varepsilon_\Lambda$ and $\varepsilon_m$ change along the cosmic time, that condition is not a coincidence but a characteristic of the de Sitter-invariant approach to cosmology.} $\gamma(t) \sim 1$, we can write
\begin{equation}
\Lambda \sim \frac{8 \pi G}{c^4} \, \varepsilon_m.
\label{54}
\end{equation}
Furthermore, as seen in Section~\ref{topo}, the critical energy density is also the same as $\varepsilon_m$ and $\varepsilon_\Lambda$. Substituting
\begin{equation}
\varepsilon_m \sim \varepsilon_c = \frac{3 H^2 c^2}{8 \pi G}
\end{equation}
in Eq.~\eqref{54}, it yields a relation between $\Lambda$ and the Hubble parameter $H$:
\begin{equation}
\Lambda \sim \frac{3 H^2}{c^2}.
\label{coscon2}
\end{equation}
Using the value $H \simeq 67.4$~Km/s/Mpc, the cosmological term is found to be
\begin{equation}
\Lambda \sim 10^{-52}~\mbox{m}^{-2},
\end{equation}
which is of the order of magnitude of the present-day observed value.\footnote{Alternatively, we can use the currently observed value of the cosmological term~$\Lambda$ to determine the Hubble parameter~$H$.} Besides providing an origin for $\Lambda$, the de Sitter-invariant approach to cosmology correctly determines its current order of magnitude. 

\subsection{The deceleration parameter}

The deceleration parameter and the Hubble parameter make up the fundamental parameters to describe the universe's evolution. It is defined as
\begin{equation}
q \equiv - \frac{\ddot{a}}{a} \frac{1}{H^2} = - \frac{\ddot{a} a}{\dot{a}^2}.
\label{q}
\end{equation}
In the ordinary case of locally Minkowski spacetimes with a cosmological constant $\Lambda$, the deceleration parameter has the form
\begin{equation}
q_M = {{\textstyle{\frac{1}{2}}}} \big(1 + 3 w \big) \Omega_m - \Omega_\Lambda.
\label{Ordiq1}
\end{equation}
On the other hand, using the Friedmann equations \eqref{dSFried1} and \eqref{dSFried2}, the deceleration parameter in locally-de Sitter spacetimes is found to be
\begin{equation}
q_{dS} = {{\textstyle{\frac{1}{2}}}} \Big[ \Omega_m \big(1 + 3 w \big) - \Omega_\Lambda \big(1 - 3 w \big) \Big].
\end{equation}
Assuming that the present-day universe's content can be fairly described by dust ($w = 0$), and using the values~\eqref{ObsOmegas}, the deceleration parameters are found to be
\begin{equation}
q_M \equiv {{\textstyle{\frac{1}{2}}}} \Omega_m - \Omega_\Lambda \simeq - 0.54
\label{Ordiq2}
\end{equation}
and
\begin{equation}
q_{dS} \equiv {{\textstyle{\frac{1}{2}}}} \big(\Omega_m - \Omega_\Lambda\big) \simeq - 0.19.
\end{equation}
One should remark that, although $q_{M}$ and $q_{dS}$ are formally and numerically similar, since $\Lambda$ is constant in the expression for $q_M$ and a function of the cosmic time in the expression for $q_{dS}$, the physical outcome of the two cases differ substantially.

\subsection{Dimensionless coupling constants}
\label{CC}

Upon including a positive cosmological term $\Lambda$ into gravitation, one adds a new repulsive interaction to spacetime physics besides the usual attractive interaction described by general relativity. Although the same field equation describes both interactions, their coupling constants differ substantially. For example, the dimensionless gravitational coupling constant for a particle of mass~$m$ is~\cite{8AdimGra}
\begin{equation}
\alpha_G \equiv \frac{m^2}{m^2_P} = {\frac{G m^2}{\hbar c}},
\label{alphag}
\end{equation}
with $m_P$ the Planck mass. According to this expression, the strength of the gravitational interaction depends on how different the particle's mass $m$ is concerning the Planck mass.
Analogously, the dimensionless coupling constant of the interaction produced by the cosmological term $\Lambda$ is
\begin{equation}
\alpha_\Lambda\equiv \frac{l_P^2}{l^2} \equiv \frac{\Lambda}{\Lambda_P} = {\frac{G \hbar}{l^2 c^3}}.
\label{alphal}
\end{equation}
According to this expression, the strength of the $\Lambda$ interaction depends on how different the particle's de Sitter pseudo-length is concerning the Planck pseudo-length. Even though both coupling constants depend on Newton's gravitational constant, their nature differs substantially. For example, whereas the gravitational coupling constant $\alpha_G$ depends on the squared mass $m^2$, the dark energy coupling constant $\alpha_\Lambda$ depends on the squared pseudolength $l^{-2}$.
For comparison, let us recall that the electromagnetic fine-structure constant for a particle with electric charge~$q$ is
defined as
\begin{equation}
{\alpha_E} \equiv \frac{q^2}{q^2_P} = \frac{q^2}{\hbar c},
\end{equation}
with $q_P = \sqrt{\hbar c}$ the Planck charge. Similarly to the other interactions, the strength of the electromagnetic interaction depends on how different the particle's electric charge $q$ is concerning the Planck charge.

\subsection{Rescuing the local conformal transformations}

Local (or proper) conformal symmetry is a broken symmetry of nature, which is expected to become an exact symmetry at the Planck scale~\cite{coleman}. However, as the Poincar\'{e}-invariant Einstein's special relativity does not include local conformal transformations in the spacetime kinematics, it is unclear how it could become relevant at the Planck scale. For this reason, local conformal transformation is sometimes considered the missing component of spacetime physics~\cite{tHooft}.

On the other hand, the de Sitter-invariant special relativity naturally includes the proper conformal transformations in the spacetime kinematics. Such in\-clu\-sion occurs because the de Sitter group is obtained from Poincar\'{e}'s by replacing translations with a combination of translations and proper conformal transformations --- known as de Sitter ``translations.''

Note that the above inclusion does not change the dimension of the spacetime's local kinematics as Poincar\'{e} and de Sitter are ten-dimensional groups. Its unique effect is to change the local transitivity of spacetime from translations to a combination of translations and proper conformal transformations. Due to the inclusion of proper conformal transformations into the spacetime's local kinematics, it is now possible to probe its role at the Planck scale~\cite{ccc}.

According to the de Sitter-invariant approach to physics, all Poincaré-invariant relativistic theories are incomplete because they lack the conformal sector brought about by the de Sitter-invariant approach. In particular, standard quantum mechanics is incomplete and must be supplemented with the conformal sector. The completeness of standard quantum mechanics has already been questioned by Einstein, Podolsky, and Rosen~\cite{EPR}, with their arguments known today as the EPR paradox. One may wonder if the de Sitter-invariant quantum mechanics, which includes the conformal sector, could somehow contribute to elucidating the EPR paradox.\footnote{This question lies outside the scope of the present paper and will be considered elsewhere.}

\section{Final remarks}

Cosmological observations in the last decades have shown that the universe's expansion is accelerating~\cite{obs1, obs2, obs3, DES}. Since general relativity does not have a solution for a universe with accelerated expansion, it is necessary to incorporate external elements into the theory to drive the observed accelerated expansion.

The $\Lambda$CDM model has two procedures for performing such inclusion. The first is to suppose the existence of a (perfect) fluid permeating the whole universe, whose energy-momentum tensor is the source of $\Lambda$. However, there are some difficulties with this procedure. To produce a repulsive effect, the fluid must satisfy an exotic equation of state not seen in any existing fluid. Furthermore, whatever is sourced by an energy-momentum current will couple to matter with the (mass-dependent) dimensionless gravitational coupling constant~\eqref{alphag}. The problem is that, due to its non-gravitational nature, dark energy should couple to matter with the conformal ($\Lambda$-dependent) dimensionless coupling constant~\eqref{alphal}. As discussed in Section~\ref{dSsource}, such a coupling can only be achieved if the proper conformal current sources dark energy. Additionally, no exotic fluid is necessary because the proper conformal current gives rise to repulsive interaction independently of the equation of state satisfied by the fluid.

The second procedure consists of adding a positive cosmological constant~$\Lambda$ to the left-hand side of Einstein's equation, which is then interpreted as a fundamental constant of nature. However, this procedure is not free of problems either. Owing to the strong equivalence principle, all solutions to the standard Einstein's equations are spacetimes that reduce locally to Minkowski. Since Minkowski has vanishing sectional curvature, one is adding a vanishing~$\Lambda$ to general relativity. This means that the usual course of adding~$\Lambda$ to the left-hand side of Einstein's equation while keeping the spacetime's local kinematics governed by the Poincaré-invariant special relativity is unjustified.

Conversely, in the de Sitter-invariant general relativity, where all solutions to the gravitational field equations are spacetimes that reduce locally to the Sitter, the cosmological term~$\Lambda$ is non-vanishing. Since it is encoded in the spacetime's local kinematics~\cite{hendrik}, it is constitutive and does not need to be added to the left-hand side of Einstein's equation. In this case, the de Sitter-invariant Einstein's equation naturally has a solution for a universe with accelerated expansion. Furthermore, as~$\Lambda$ does not appear explicitly in the gravitational field equation, it is no longer required to be constant by the second Bianchi identity. This theory allows an entirely new view of the universe and offers new tools to explore it. 

There is a growing feeling today that solving the current problems of quantum gravity and cosmology will require new physics. By construction, the de Sitter-invariant approach to cosmology gives rise to deviations concerning the Poincaré-invariant approach, precisely for energies comparable to the Planck energy and at the universe's large scale. Accordingly, this approach may eventually provide the necessary new physics to tackle those problems.

\section*{Acknowledgments}

DFL thanks Dalhousie University, Canada, for a Ph.D. scholarship. JGP thanks Conselho Nacional de Desenvolvimento Científico e Tecnológico, Brazil, for a research grant (Contract 312094/2021-3). JRS thanks Conselho Nacional de Desenvolvimento Científico e Tecnológico, Brazil, for a Ph.D. scholarship (Contract 166193/2018-6).


\end{document}